# Making Characters Count. A Computational Approach to Scribal Profiling in 14th-Century Middle Dutch Manuscripts from the Carthusian Monastery of Herne[1,2]

## Introduction

### Herne

The cultural significance of the Herne charterhouse in nowadays Belgium cannot be overstated: no other mediaeval scribal production centre in the Low Countries can match this monastery's voluminous output. A total of 54 quires and full manuscripts have been preserved from this monastery, with 43 of these written in the Dutch vernacular. This number is particularly impressive considering that most of these works were copied in the relatively brief period between 1350 and 1400 (Kwakkel 2002a). Additionally, many salient works that are still extant today are *only* known because they were copied in Herne, such as Filip Utenbroeke's sizable contribution to the *Spiegel historiael*, the Middle Dutch adaptation of the *Speculum historiale*. In short, during the latter half of the 14th century, the Herne charterhouse quickly established itself as an influential centre for the production, and dissemination of vernacular literature in the Low Countries.

The monastery of Herne, which is located approximately thirty kilometres southwest of Brussels, was the first Carthusian charterhouse to be established in the Low Countries (1314). Carthusian monasteries like that of Herne distinguished themselves from other monasteries on various fronts; all of which stemming from their isolated way of living and the challenging set of customs that they entertained. These customs – or *Consuetudines* – established by Guigo I (1083-1136), the fifth prior of the Grand Chartreuse, advocate a life of profound silence, permitting Carthusians to speak only at rare occasions and for limited periods of time (Gaens and Grauwe 2006). Moreover, since Carthusians were bound by these rules not to leave the monastery for the purpose of preaching, the task of copying books assumed a vital role in their spiritual outreach. As Guigo's rule dictates, books should be held in equal esteem to the spoken word, rendering the transcription and dissemination of spiritual texts the sole apostolate they could partake in.[3] This dedication is reflected in their daily routine, as monks would typically devote several hours each day to meticulous copying in the silent solitude of their own, isolated cells.

---

[2] Authors: Caroline Vandyck (University of Antwerp, caroline.vandyck@uantwerpen.be), Wouter Haverals (Princeton University) & Mike Kestemont (University of Antwerp).

Furthermore, Guigo emphasises the importance of faithful and correct transcriptions in his *Consuetudines*: producing error-free manuscripts was a priority for the Carthusians. To maintain their high-quality standards, the Carthusians engaged in intense collaborative efforts. However, since verbal communication was strictly limited and since works were copied in isolation, the scribes resorted to more creative strategies to facilitate these collaborations. Through marginal annotations, for instance, in which they discussed their joint projects, they coordinated and streamlined the process of cooperation. This allows us, present-day scholars, for a unique glimpse over the shoulders of these scribes as they "discuss" their work, often focusing on the quality of translations (Fig. 1). Such marginal notes are rare and therefore of great value for the study of scribal practices, since they provide a rare insight into the charterhouse, its scribes, and their dynamic.

The silent yet collective approach to manuscript production sets Herne apart from other monasteries belonging to other religious orders. Another unique feature of Herne's monastic life is the ubiquity of so-called "doublets" (*doubletten*) – multiple copies of the same text – within their library (Kwakkel 2002a, 47–52). The absence of a communal scriptorium or reading room compelled the Carthusians to work with manuscripts in the isolation of their cells. Consequently, the monastery needed to host several copies of popular works, enabling monks to access and copy these texts concurrently. In Kwakkel's reconstruction of the Herne library, examples of such doublets are Vienna, ÖNB, Cod. 12.857 and Saint Petersburg, BAN, O 256 which both contain evangeliaries (2002a, 189).

Furthermore, the Carthusians' manuscript work extended beyond adding to their own library; they also accepted commissions – referred to as "pro pretio" (Kwakkel 2003, 202) – for urban patrons, merchants, and even other religious orders, an endeavour that was notably uncommon. The quality of such books could vary significantly depending on their destined readers. Manuscripts intended for internal use were typically plainer, in contrast to those crafted for external circulation, which were often more polished and accessible. Adapting to the quality demands of the project at hand, one Herne scribe (β) even employed a range of handwriting styles, from "low" to "medium" to "high".

---

[3] "Libros quippe tanquam sempiternum animarum nostrarum cibum cautissime custodiri et studiosissime volumus fieri, ut quia ore non possumus, dei verbum manibus predicemus" (Guigues 2001). Our translation: "For indeed, we wish books, as the perpetual nourishment of our souls, to be guarded most carefully and to be made with the utmost diligence, so that, because we cannot [preach] by mouth, we may proclaim the word of God with our hands."

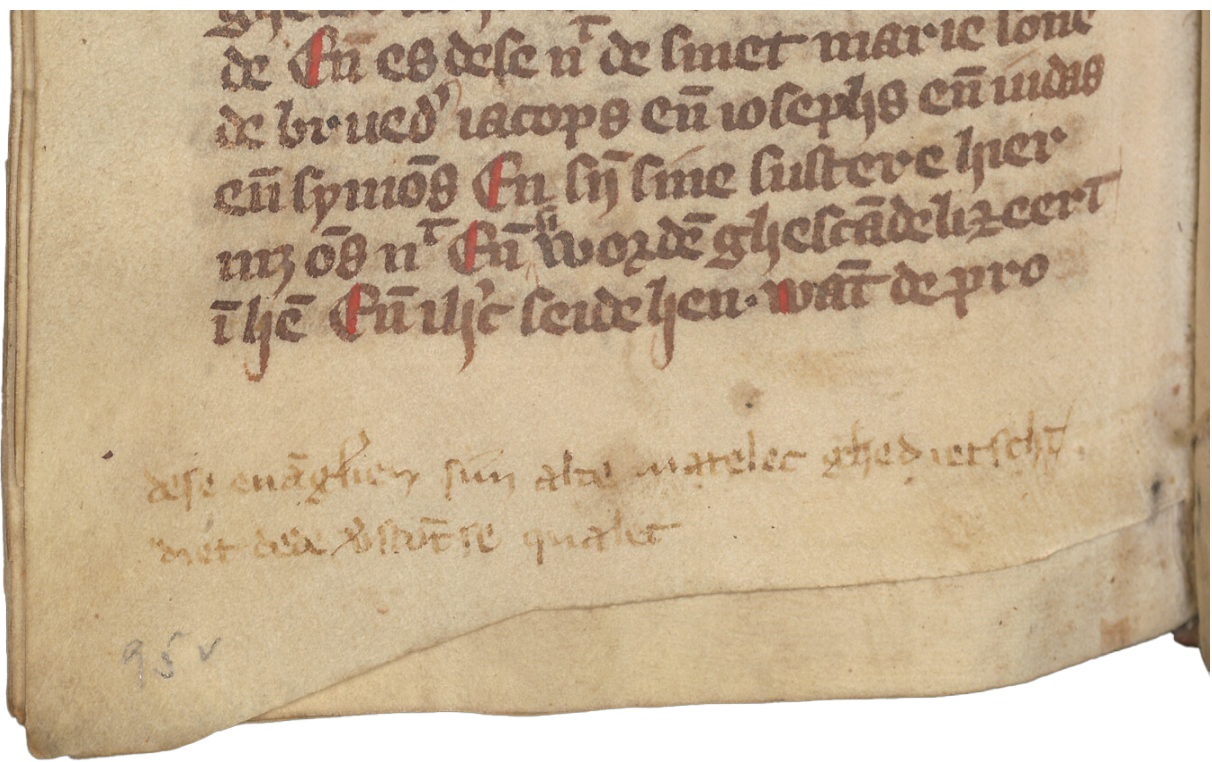

Fig. 1. The footer text in a late fourteenth-century Middle Dutch gospel book (Vienna, ÖNB, Cod. 12.857, 95v, modern foliation) reveals a Carthusian monk's frustration with the quality of translation, candidly noting, "These Gospels have been translated very poorly into the vernacular; [the person] who did it understood them inadequately," see also Kwakkel 2018. The primary text is copied by scribe α, the comment was later added by scribe β.

Another distinguishing feature of the Herne monastery is the significant impact it had on the literary field and knowledge transfer in the Low Countries through their dissemination of manuscripts. Herne was renowned not just as a hub for manuscript production through copying but also as a *translatorium*, a centre of translation (Kwakkel 2002a). A prime testament to this is the ambitious project undertaken by the anonymous "Bible Translator of 1360" (*Bijbelvertaler van 1360*), whose *nom de plume* derives from his most ambitious project: the translation of the historical books of the Old Testament. The vernacularization of the text was a risky enterprise: the Bible Translator of 1360 embarked on this colossal task, exposing himself to the scrutiny of conservative theologians who believed that the Bible should not (or even could not) be translated into the vernacular. He translated eleven other works as well, at a time when the translation of religious texts from Latin was still controversial.

In the late fourteenth century, seventeen people lived in the charterhouse of Herne (Kwakkel 2002a, 85). Based on formal aspects, Kwakkel was able to distinguish thirteen separate hands who were active between 1350 and 1400. This indicates that the majority of charterhouse residents was involved in book production activities at the time. Presumably, individual residents focused on one specific task, such as translation, book binding, or copying (Kwakkel 2002a). This specialisation, coupled with collaborative efforts - where two monks could divide their writing, correct each other, or one could copy while the other added rubrication - facilitated quick and error-free translations. Accordingly, the scribes devoted much attention to the quality of translations and adopted a unique way of signalling the need for corrections. In the margins of several manuscripts, a struck-out -*d* or delta can be found (Fig. 2). Kwakkel presumes this is a continental adaptation of the insular practice of correcting omissions with the struck-out letters h (for *haec*) and d (for *deorsum* or *deletum*) (2003, 200). The correction delta, other paleographic aspects and collaborations allowed Kwakkel to identify and situate the thirteen different scribes in Herne. The two most productive scribes are the so-called "Speculum Scribe" (*Tweede Partie-kopiist*), named after his substantial copy of the "Second Part" (*Tweede Partie*) of the *Spiegel historiael*, and the "Necrology" scribe, who wrote the necrology of the charterhouse. The former copied seven manuscripts in total, corrected two codices and two production units, and the latter copied eight manuscripts, with the Speculum scribe present in four of them. This indicates that the two worked closely together (Kwakkel even argues that Brussels, KBR, 2849-51 went back and forth no less than fifteen times between the two scribes) and most likely lived in the same environment. Since the Necrology scribe updated the Herne-necrology, that environment is most likely the charterhouse of Herne. Accordingly, all other scribes which collaborated with him, could also be located in Herne (Kwakkel, 2002a).

Finally, it deserves emphasis that the Herne scribes consistently wrote without revealing their identities; not one of the surviving works includes a scribe's self-attribution. Consequently, some of the current scribal attributions are controversial. Whereas previous research regarding Herne often tried to identify the historical figures behind the hands based on paleography and codicology, our approach concentrates on the linguistic characteristics of the scribes. Accordingly, in this study we explore the Herne dataset for the first time using computational methods from the field of scribal modelling.[4] Contrary to authorship attribution, scribal modelling seeks to identify scribes rather than authors. The structure of our study is laid out as follows: after an overview of scribal modelling, a summary of the used materials is included. We then examine the extent to which abbreviations can inform us about the scribes' identities. Finally, we compare these findings with a stylometric analysis of the various Herne

---

hands.[5] This analysis demonstrates that the minutiae of a writing system can be instrumental in shedding light on the identities of mediaeval scribes.

---

 Code available on https://github.com/Caroline-Vandyck/making-characters-count

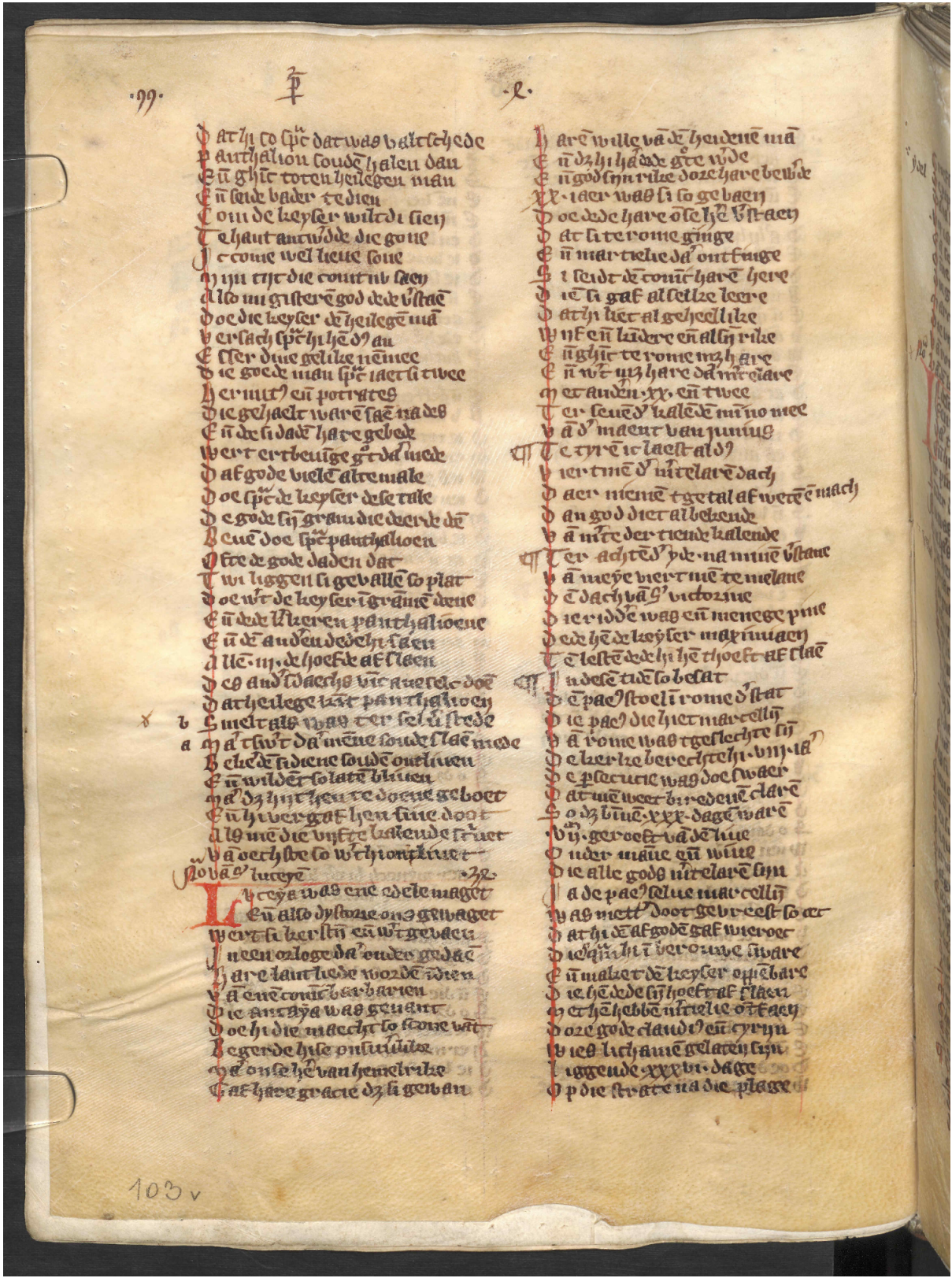

Fig. 2. An example of a correction delta in column a of manuscript Vienna, ÖNB, Cod. 13.708, 103v (modern foliation). The correction delta indicates that verse *a* and *b* need to be switched.

## Scribal Modelling

Few authors of Middle Dutch literature are known today, with the majority of manuscripts being transmitted anonymously. Consequently, it is not surprising that a major part of mediaeval literature studies is concerned with identifying authors and their oeuvre (either traditionally or computationally, e.g., van Mierlo 1929b; Lievens 1960; van Dalen-Oskam 2007; Kestemont 2018). In these studies, particularly the computational analyses, researchers make great effort to overcome scribal variation when trying to identify the original author of a text. Since Middle Dutch did not have a standard language variety or standardised spelling, mediaeval spelling was highly phonological and allowed for scribes' "personal" preferences to seep through in the texts they copied (McIntosh 1975; Kestemont, Daelemans, and De Pauw 2010). Accordingly, spelling displays "dialectal pronunciations and local spelling habits" (Kestemont, de Pauw & Daelemans, 2010, 287). Subsequently, even the most frequent words could be written in different ways, sometimes even by the same scribe (Kestemont, de Pauw & Daelemans, 2010; Thaisen 2010). Apart from the lack of a supra-regional language variety, scribes could also make mistakes, or manipulate texts on purpose, based on the presumed preferences of their audience (Thaisen 2010). All this scribal variation makes authorship attribution a complicated, yet interesting task in which spelling and scribal variation are often seen as hurdles to overcome. However, what if we were to zoom in on the variation and the scribes themselves? The variation they produce also forms an interesting research topic (e.g. McIntosh 1975; van Dalen-Oskam 2007; Kestemont 2015; Kestemont and van Dalen-Oskam 2009; Haverals and Kestemont 2023). It can provide more information on dialects, the identity of scribes, their dynamics or influence on each other, and the evolution of their writing process. The study of this is called "scribal modelling" or "profiling".[6] Scribal modelling was first defined by Angus McIntosh in 1975. Parallel to the hypothesis that all authors possess a unique fingerprint, he argues that scribes leave their personal stamp on a copied text as well.

McIntosh set out to create scribal profiles in Middle English texts (1975). The underlying principle is simple: distinct profiles would point to different scribes, while identical ones would suggest a single scribe. The scribal profiles consist of two different classes of features that complement each other, namely "linguistic" and "graphetic" ones (222). The linguistic features refer to data concerning "the lexical and morphological (and perhaps syntactical) habits of the scribe when writing, but also about his graphological system" (222). The graphetic features allude to those aspects that normally a "paleographer primarily concerns himself [with]" (223). On the topic of the linguistic profile, McIntosh states that most of the scribes are inclined to use a "linguistic form which closely reflect[s] his personal written language characteristics" (224). Therefore, even scribes operating in the same period and setting, such as Herne between 1350-1400, are not likely to exhibit identical writing habits. However, McIntosh also noted that two scribes might present strikingly similar linguistic profiles, appearing as if they were the same individual, even in the absence of a direct scribal link. This phenomenon arises, for example, when a scribe closely follows an exemplar. In such instances, an additional analysis of paleographic features could provide further clarity on the scribes' identities. Our study primarily concentrates on the linguistic aspect, especially the orthography of manuscripts, which lies at the intersection of language and script. As such, our research aims to supplement Kwakkel's earlier work, which was mainly 'graphetic' (focusing on paleography and codicology). Consequently, we consistently cross-reference our linguistic findings with Kwakkel's paleographic insights (2002a).

---

[6] *Scriptométrie* in French.

## Materials

The corpus that we use is outlined in Haverals and Kestemont (2023b). It is near-complete, encompassing the majority of Middle Dutch manuscripts linked to Herne during the reference period 1350-1400, as per Kwakkel's library reconstruction (2002a). In particular, this means that the manuscripts were either produced or corrected in Herne, or that they were at least housed there at some point. Additionally, we have included one more manuscript: Ghent, UL, 941 (Hadewijch ms. C). Although not originating from Herne, it bears a close relationship in content to Brussels, KBR, 2877-79 and Brussels, KBR, 2879-80 (Hadewijch ms. B and A), as discussed in Kestemont (2015) and Haverals & Kestemont (2023a). Furthermore, non-vernacular (e.g. in Latin) texts as well as short texts were excluded from the corpus.[7] A summary of the codices, their contents and characteristics can be found in Table 1.

The corpus contains "hyper" diplomatic transcriptions of the included manuscripts which are freely available on Zenodo.[8] The process of creating this corpus started with the acquisition of photographic facsimiles of each manuscript.[9] A portion of these manuscripts was then transcribed manually, adhering to a "graphematic" transcription practice as outlined by Stutzmann (2011). This approach entails reducing each letter form to a standardised representation while preserving the original spelling of the text and maintaining all abbreviations. However, a distinction was made between 'u'/'v' and 'i'/'j' spellings, as this is traditionally done in Middle Dutch studies because these variations in some cases mark distinct phonetic realisations. Nevertheless, the transcription does not differentiate between allographic variations such as the long s (ſ) and the lowercase s, or the r rotunda (ꝛ) and the lowercase r.

---

[7] Short texts such as the book list "die dietsche boeke die ons toebehoeren"; see Kwakkel, 2003, pp. 24-30.
[8] https://zenodo.org/records/10005366
[9] Which are available on request: https://zenodo.org/records/10010382

Building on this manual transcription effort, and incorporating several editions already available in digital format, we developed a Handwritten Text Recognition (HTR) model using Transkribus. The model was trained on 1,331 folios and obtained a Character Error Rate (CER) of 2.7%, meaning that for every 100 characters, less than three are transcribed incorrectly. Subsequently, the remaining 4,828 folios were transcribed using this HTR model. The primary aim of this model was to produce transcriptions that digitally replicate the manuscripts with the highest possible fidelity to the original sources. This digital replication was conducted in alignment with the standards of editorial practice outlined above, ensuring an accurate and faithful rendition of the original texts (Fig. 3).

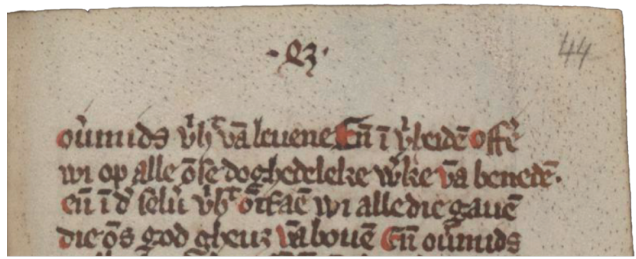

.43.
ou'mids vⁱhᵗ vā leuene Eñ ī vⁱheidē off'e
wi op alle ōse doghedeleke w'ke vā benedē .
eñ ī d' selu' vⁱhᵗ ōtfaē wi alle die gauē
die ōs god gheuꝛ vā bouē Eñ ou'mids

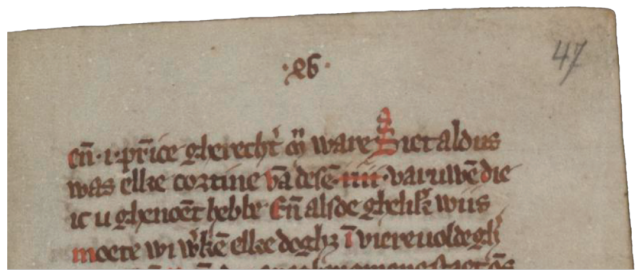

.46 .
eñ .i. prīce gherecht' mⁿ ware Siet aldus
was elke cortine vā desē . mi . varuwē die
ic u ghenoēt hebbe . Eñ alsoe ghelik'wiis
moete wi w'kē elle doghꝛ ī viereuoldegh'

Fig. 3. (top) An excerpt of a page of manuscript Brussels, KBR, 3091, 44r (modern foliation) and its manual transcription. (bottom) An excerpt of a page of manuscript Brussels, KBR, 3091, 47r (modern foliation) and its automatic HTR transcription. An error can be found line 3, where *mi* should be *iiii*.

Finally, an important terminological clarification is necessary regarding the corpus used in our study. In his analysis, Kwakkel (2002a) frequently refers to "production units" instead of entire manuscripts, which typically consist of multiple distinct codicological units. Production units are defined as "groups of quires that formed a material unity at the time of production. Such quires were copied 'in one go', by either one or more scribes. A codex may contain several production units" (2002b, 13). These units could be bound together at various stages, sometimes even in post-medieval times. Consequently, production units can generally be dated more precisely than entire (composite) manuscripts. An illustrative example is Vienna, ÖNB, Cod. 13.708, which is composed of 11 production units. These are further classified into an 'old core' (consisting of units II, X, and XII, written in 1393-1394) and a 'new core' (comprising the remaining production units, added in 1402), as per Kwakkel (2002a).

Table 2 reveals that some scribes contributed significantly more than others (both in terms of concrete production units and overall character counts), suggesting a high degree of specialisation within Carthusian manuscript production. While some scribes focused more on writing, others presumably concentrated on correcting or preparing parchment. The character counts presented in Table 2 have been adjusted after cleaning the text of any artefacts resulting from the HTR model and the removal of punctuation. Punctuation was excluded for three key reasons: firstly, the scribes themselves often omitted it; secondly, the HTR model might misinterpret imperfections in the parchment as punctuation marks; and thirdly, correctors occasionally employed punctuation as markers for required text additions or modifications.

| Codex | Scribe(s) | Middle Dutch Pages | Main Content | Words | Unique Words |
|---|---|---|---|---|---|
| BRUSSELS, KBR, 1805-1808 | α | 84 | Gregory the Great (dialogues) | 61,332 | 9,057 |
| | β | 7 | Hagiographies; sermons | 5,558 | 2,219 |
| | Other | 41 | Gregory the Great (dialogues) | 28,613 | 5,931 |
| BRUSSELS, KBR, 2485 | β | 135 | Rule of Saint Benedict | 27,442 | 4,744 |
| BRUSSELS, KBR, 2849-51 | α | 693 | New Testament | 154,399 | 19,718 |
| BRUSSELS, KBR, 2877-78 | Other | 330 | Hadewijch (ms. B) | 90,181 | 12,481 |
| BRUSSELS, KBR, 2879-80 | Other | 202 | Hadewijch (ms. A) | 82,113 | 15,267 |
| BRUSSELS, KBR, 2905-09 | α | 200 | *Hore dochter* | 46,954 | 8,418 |
| | Other | 13 | calendar | 1,078 | 603 |
| BRUSSELS, KBR, 2979 | α | 1 | Gospel of Matthew | 216 | 137 |
| | Other | 333 | Gospels | 84,558 | 18,232 |
| BRUSSELS, KBR, 3091 | ε | 451 | John of Ruusbroec | 75,155 | 9,943 |
| | Other | 1 | Note | 24 | 23 |
| BRUSSELS, KBR, 3093-95 | α | 300 | *Manuale*; pseuso-Bernardus; Bonaventura | 33,661 | 6,804 |
| | β | 72 | *Der minnen gaert* | 6,751 | 2,151 |
| | Other | 2 | Note; Table of contents | 32 | 31 |
| BRUSSELS, KBR, 394-98 | Other | 68 | Rule of Saint Benedict | 25,775 | 5,177 |
| GHENT, UB, 1374 | α | 264 | *Spiegel historiael; Martijns* | 77,717 | 11,838 |
| GHENT, UB, 941 | Other | 182 | Hadewijch (ms. C) | 89,585 | 14,892 |
| PARIS, BIBLIOTHÈQUE MAZARINE, 920 | γ(?) | 53 | Hugh of Saint Victor | 11,859 | 2,681 |
| | Other | 232 | John of Ruusbroec; Hadewijch | 43,034 | 12,091 |
| PARIS, BIBLIOTHÈQUE DE L'ARSENAL, 8224 | Other | 328 | Heinrich Seuse's *Horologium* | 76,633 | 17,988 |
| SAINT PETERSBURG, BAN, O 256 | Other | 416 | Gospels | 70,950 | 23,654 |
| VIENNA, ÖNB, 12.857 | α | 459 | Gospels | 84,063 | 11,352 |
| | Other | 20 | | 2,334 | 1,231 |
| VIENNA, ÖNB, 12.905 | Other | 400 | Epistle pericopes | 73,196 | 12,330 |
| VIENNA, ÖNB, 13.708 | α | 490 | Gerard Zerbolt of Zutphen; *Spiegel historiael*; John of Ruusbroec | 281,228 | 36,759 |
| VIENNA, ÖNB, 65 | α(*) | 86 | Heinrich Seuse's *Horologium* | 35,764 | 7,051 |
| | γ | 100 | Heinrich Seuse's *Horologium* (continuation) | 38,396 | 7,267 |

Table 1. Summary of codices and scribal contributions in Middle Dutch manuscripts originating from Herne, as identified by Kwakkel (2002a). This table showcases the contributions of the main scribes α, β, γ, and ε across various codices within the corpus, emphasising that these were the recurrent hands. The table provides details on the number of unique folio sides containing text in Middle Dutch content ('Middle Dutch Pages'), total word count ('Words'), and the count of unique words ('Unique Words') for those pages. Notably, Ghent, UL, 941, Paris, Bibliothèque de l'Arsenal, 8224, and Saint Petersburg, BAN, O 256 while not written in Herne, are included due to their close association with the monastery. Vienna, ÖNB, 12.905 is included since Kwakkel (2002a) marks the possibility that it was written in Herne. An * is assigned to scribe α in Vienna, ÖNB, SN 65 to denote that subsequent analysis suggests this may not be α's hand. Additionally, a question mark accompanies γ in Paris, Bibliothèque Mazarine, 920, indicating Kwakkel's uncertainty about this particular hand.

| Scribe | Codices | Production Units | Characters (no grapheme) | Characters (grapheme) | 5,000 Character Segments | Unique Characters | TTR |
|--------|---------|------------------|--------------------------|------------------------|--------------------------|-------------------|-----|
| $\alpha$ | 9 | 32 | 3,609,770 | 3,608,346 | 706 | 107 | 0.22 |
| $\beta$ | 3 | 4 | 194,756 | 194,722 | 36 | 80 | 0.24 |
| $\gamma$ | 1 | 3 | 194,809 | 194,806 | 37 | 73 | 0.22 |
| $\gamma$? | 1 | 1 | 58,871 | 58,869 | 11 | 69 | 0.23 |
| $\epsilon$ | 1 | 1 | 365,975 | 365,837 | 73 | 75 | 0.13 |

Table 2. Summary containing statistics about the recurring Herne scribes. From left to right: the amount of codices their hand can be found in (only if the scribe contributed at least one full page of main text in Middle Dutch to the manuscript), the number of production units they produced, the amount of total characters they have produced (after cleaning) - without and with applying the grapheme function (cfr. infra), the amount of segments of 5,000 characters (including whitespace, with the grapheme function), the amount of unique characters they employ, and their type-token ratio (in words) – both with the grapheme function.

## Abbreviations

Our first examination of the corpus focuses on the scribes' use of abbreviations, which is a central aspect of their copying practice. We determine the "abbreviation density" for each scribe by calculating the proportion of characters they abbreviated. In this sense, "abbreviation density" is defined as the ratio of the total number of brevigraphs (glyphs representing two or more characters (Honkapohja 2021)) to the total number of letter characters in a text.[10] While this concept may seem straightforward, its computation is more complex than it appears. Accurate calculation requires graphematic reproductions of the manuscripts, as described above, and also outlined by Robinson and Solopova (1993), which involves closely replicating the original manuscript's spelling, including brevigraphs, letters, and other glyphs. To understand why even then the calculation is not as straightforward as it seems, some technical background information is necessary.

Text processing by computers requires converting glyphs into numerical data through 'text encoding'. The Unicode Standard (*Universal Coded Character Set*) is a widely used system for this purpose, designed to map all known glyphs to unique 'code points'. In Western languages, typical glyphs such as individual letters, are usually linked to a single code point. However, combining characters like accents or macrons are allocated separate code points. When such a character follows a main character, for instance an acute accent following an 'e', it is intended to merge with the main character.[11] Glyphs formed from two separate code points are identified as decomposed characters (U+0065 and U+0301). Conversely, Unicode encompasses precomposed characters as well, representing frequently used combinations of main and combining characters with distinct, single code points. Hence, 'é' might be encoded as a singular precomposed character (U+00E9) or as a decomposed combination of 'e' and an acute accent (U+0065 and U+0301). The visual representation of a glyph does not indicate whether it is precomposed or decomposed. This distinction critically impacts the computational measurement of text length: a precomposed character is counted as one, while a decomposed character is counted as multiple characters. As a result, manuscripts abundant in abbreviations and brevigraphs could, counterintuitively, exhibit a higher character count compared to manuscripts where words are written out in full.

---

[10] Abbreviation densities based on the ratio of abbreviated words to full-length words were also calculated.

[11] https://unicode.org/standard/standard.html

In manually transcribing the materials for our corpus, we opted to utilise precomposed Unicode tokens as much as possible, specifically those recommended by the Medieval Unicode Font Initiative (MUFI).[12] However, potential inconsistencies may still be present in both manually-created and HTR-generated transcriptions. Furthermore, not every glyph can be reduced to a single Unicode point (for example, an 'r' combined with a macron exists only as a decomposed character sequence). The Python package *grapheme* assists in accurately interpreting these characters.[13] As part of its functionality, it can recognize and treat glyphs composed of multiple Unicode points as single characters. As a result, character counts derived using *grapheme* correspond to what would be visually perceived by a human observer. For instance, consider the Middle Dutch token "stoer̄", employed by scribe α to abbreviate the word "stoerm". Even though 'r̄' is a decomposed character made up of two Unicode points, the grapheme library interprets 'r̄' as one glyph. This interpretation ensures the alignment of our digital transcriptions with the original manuscript's visual representation. Character counts for each scribe, calculated both with and without the use of *grapheme*, are presented in Table 2. This underlines an often-neglected aspect of digital encoding in mediaeval manuscripts, where there is ambiguity about whether brevigraphs are represented by single or multiple Unicode tokens.

After confirming the accuracy of the character counts to reflect the original manuscripts, we calculated the average abbreviation density for each scribe, as depicted in Figure 4. The graph presents the main scribes from the Herne corpus, identified by Erik Kwakkel as α (the Speculum scribe), β (the Necrology scribe), ε, γ, and γ(?) – all highlighted in a darker shade for distinction. The figure also includes abbreviation densities for other scribes whose work was once part of the Herne library, though their affiliation with the monastery itself is either uncertain or not established. The data illustrates that the Herne-associated scribes generally used abbreviations more frequently than the 'unknown' scribes. For instance, scribe ε's abbreviation density at the character level is notably high at 0.18, indicating that nearly one in

---

every five characters he wrote down marks an abbreviation. When considering whole words, this density suggests that every second word was abbreviated. Such a high frequency of abbreviation is typical in mediaeval Latin texts, but it is exceptionally high for vernacular languages (Honkapohja 2021).

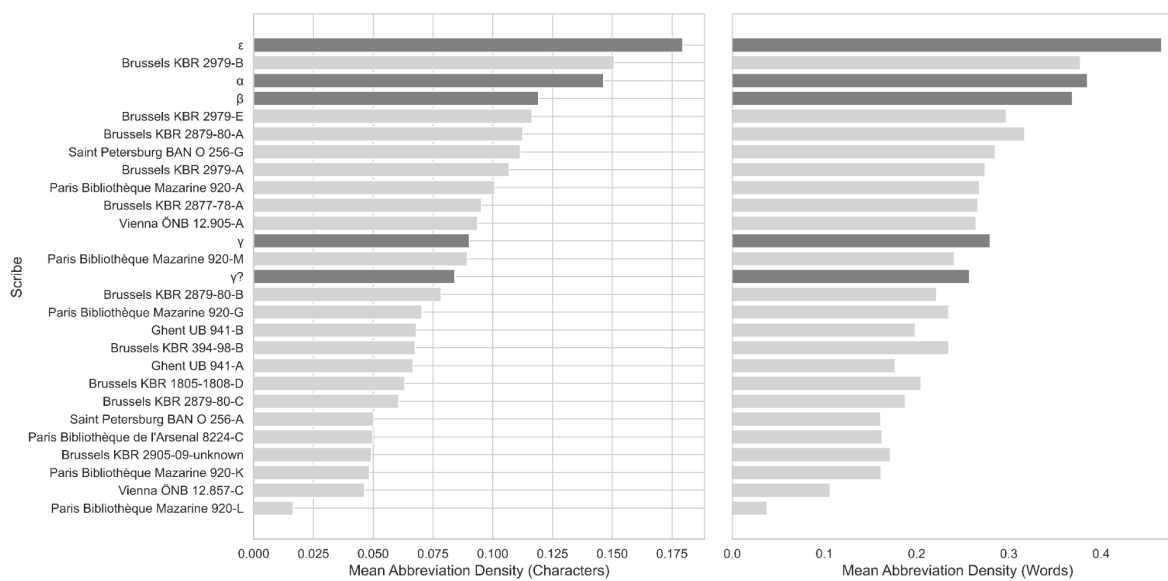

Fig. 4 Mean abbreviation densities by scribe, with Herne scribes distinguished in dark shading and 'unknown' scribes (of uncertain origin) in light shading. The label 'γ?' indicates uncertainty about the identification of this scribe as γ, as noted by Kwakkel (2002a). Vienna, ÖNB, SN 65 (production unit II) is included in α's works (cfr. infra). The left side of the graph shows the mean abbreviation density measured at the character level, whereas the right side presents the mean abbreviation density at the word level.

Our investigation into the abbreviation practices of individual scribes brought scribe α's work into focus. We computed the mean abbreviation density for all the codices and production units attributed to him, illustrated respectively in Figures 5 and 6. Notably, the manuscript Vienna, ÖNB, SN 65, a collaborative effort between scribes α and γ, shows an abbreviation density for α's sections significantly lower than that found in his other manuscripts. This difference is also evident from a 5,000-character sample-based examination, which highlights the reduced abbreviation density in Vienna, ÖNB, SN 65 (production unit II) compared to the rest of α's corpus (Fig. 7, top). Similarly, Vienna, ÖNB, Cod. 13.708 (Fig. 7, bottom) and Vienna, ÖNB, Cod. 12.857 (Fig. 7, middle) demonstrate lower abbreviation densities relative to α's other works, albeit to more subtle extents.

Fig. 5 Mean abbreviation density (character-level) for scribe α by manuscript signature.

Fig. 6 Mean abbreviation density (character-level) for scribe α within production units, detailing differences between the 'old core' and 'new core' of Vienna, ÖNB, Cod. 13.807.

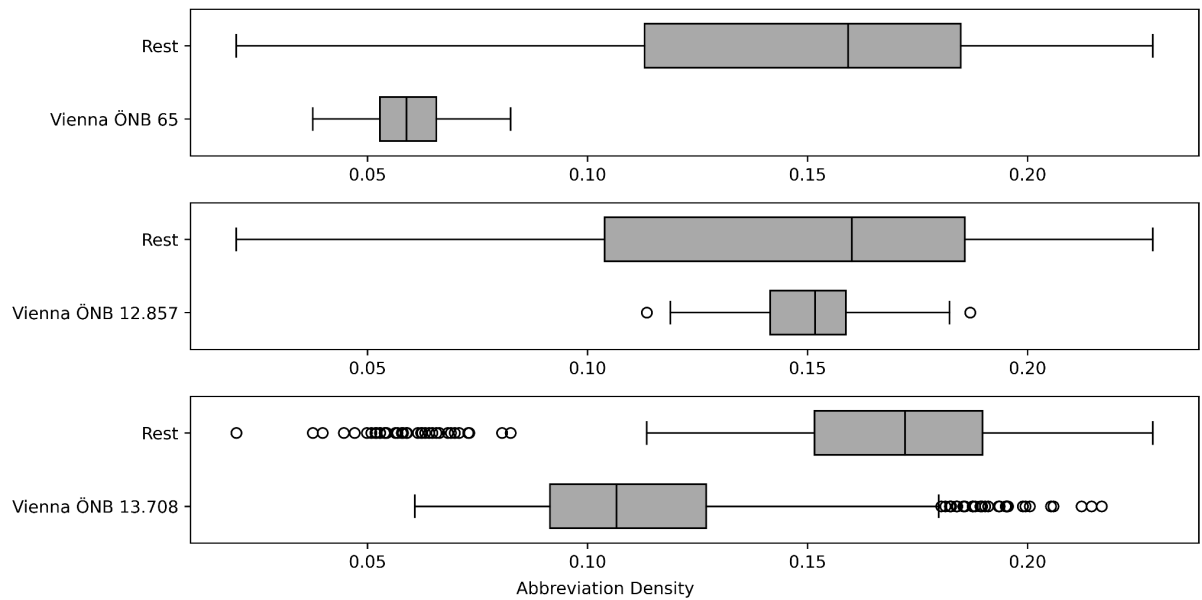

Fig. 7 Comparative analysis of scribe α's abbreviation densities across selected manuscripts. This visualisation presents boxplot comparisons of the abbreviation densities at the character-level for scribe α's work in 5,000-character samples taken from three notable manuscripts – Vienna, ÖNB, SN 65 (top), Vienna, ÖNB, Cod. 12.857 (middle), and Vienna, ÖNB, Cod. 13.708 (bottom) – against the aggregate data from the rest of his corpus.

In examining Vienna, ÖNB, Cod. 13.807, an intriguing pattern emerges concerning the abbreviation density across its production units. As detailed previously, this codex is segmented into eleven production units that Kwakkel categorises into two distinct cores. The "old core" comprises three production units (II, X, and XI) and features 56 various texts and excerpts translated from Latin to Middle Dutch. In contrast, the "new core" spans eight production units (I and III-IX) and predominantly contains two texts, with the first unit presenting Gerard Zerbolt of Zutphen's *Super modo vivendi* in Middle Dutch and units III-XI encompassing Philip Utenbroeke's *Spiegel historiael*. While the old core was penned in 1393-1394, the new core was appended in 1402, according to Kwakkel (2002a). Interestingly, the old core exhibits a denser use of abbreviations than the newer segments (Fig. 6). However, an exception is observed in production unit I, which does not conform to this pattern. This anomaly is particularly intriguing given Biemans's assertion (1997, 117–19) that production unit I might have been included in the codex as early as 1394. This discrepancy prompts a reevaluation of the divergence in abbreviation densities. However, Kwakkel refuted that thought since the foliation on the last page of production unit I was only added after the two cores were bound together. Accordingly, we might have to look in a different direction to find the source of this deviation in abbreviation densities. In that regard, it is interesting to note that production units I, II, X, and XI were translated from Latin to Middle Dutch by scribe α himself (Kwakkel 2002a, 130, footnote). This raises the possibility that α had a greater tendency to abbreviate in his own translations than in his transcription of others' work. Such a hypothesis presents a compelling avenue for future investigation, potentially shedding more light on α's scribal practices.

Finally, upon examining the abbreviation densities within individual codices, another pattern emerges: codices presumably created for personal use exhibit a higher use of abbreviations compared to those intended for an external audience. This phenomenon was examined through the abbreviation densities found in two versions of the same didactic poem included in our corpus: Jacob van Maerlant's *Derde Martijn*, both transcribed by scribe α. The first version is contained in Vienna, ÖNB, Cod. 13.708, and the second in Ghent, UL, 1374.[14] Although the content of the manuscripts is nearly identical, their layouts are quite different, as illustrated in Figure 8. The Viennese manuscript is meticulously formatted, with each verse line placed neatly below the previous one, and the text arranged into two columns. In contrast, the Ghent manuscript presents the verses in a prose-like, continuous flow and is characterised by lower quality parchment, resulting in a less formal presentation. These distinctions imply that the Ghent manuscript might have served as a personal copy for the scribe (Kestemont, forthcoming). Therefore, these manuscripts, each containing parallel versions of the *Derde Martijn*, provide an ideal comparison to assess the impact of the intended audience on abbreviation practices. As shown in Figure 9, the Ghent manuscript indeed has a significantly higher abbreviation density – almost twice that of the Viennese manuscript, with mean abbreviation densities of 0.147 and 0.085 at the character level, respectively. This observation reinforces the idea that manuscripts crafted for a scribe's personal reference are inclined to feature a more condensed use of abbreviations compared to those prepared for wider circulation.

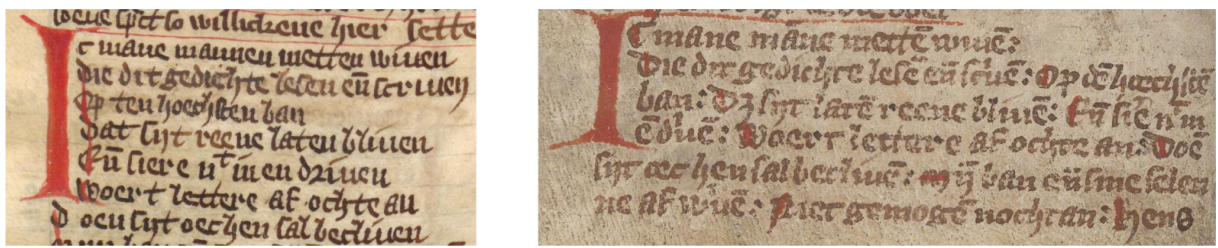

Fig 8. The opening verse lines of the *Derde Martijn* in Ghent, UL, 1374, 113v (left) and Vienna, ÖNB, Cod. 13.708, 218r (to the right), both following modern foliation. The Ghent manuscript features a prose-like structure with verses separated by colons.

---

14   A prior analysis of Ghent, UL, 1374 can be found in the introduction of the diplomatic edition of the manuscript by Gabriël and Kestemont (2023).

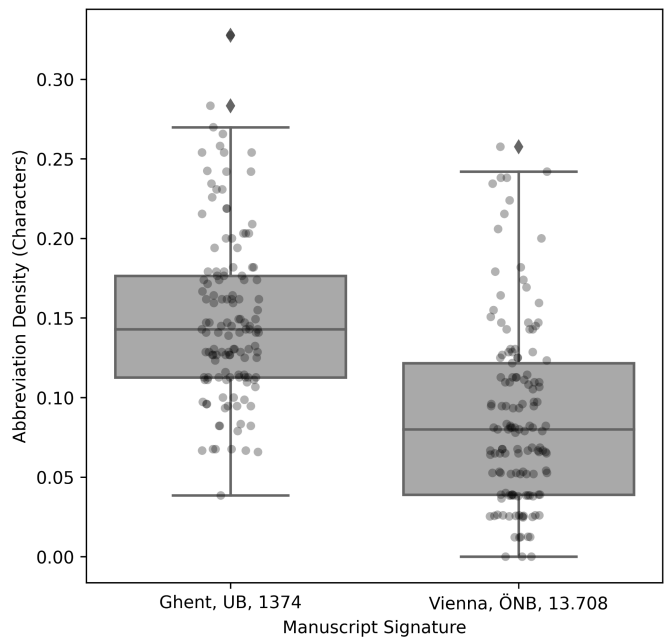

Fig. 9. The distribution (box plot and strip plot) of the use of abbreviations in Ghent, UL, 1374 as well as Vienna, ÖNB, Cod. 13.708.

# Stylometric Analysis

## Bags-of-Characters

In concluding our exploration of the copying styles from the Herne scribes, we extend our analysis to a stylometric examination. This method seeks to investigate how the different production units and scribes relate to each other in terms of writing (or in our case: copying) style. Such an inquiry necessitates a computational model of the texts, a numerical representation that quantifies the textual data. The "bag-of-words" model (BOW) is a prevalent method for this purpose. This model disregards the original sequence of words, losing all context (hence referring to it as a "bag"). In essence, each row of this table aligns with a text in the corpus, and each column signifies a distinct word. Frequencies of these words populate the table, providing a count for each text (Sebastiani 2002, 10-12). Most often, not the absolute, but the relative frequencies are calculated: the total number of a word's occurrences divided by the total number of words in the document. Another approach is to calculate the "term frequency-inverse document frequency" (TF-IDF), which assigns greater significance to words that are uncommon overall but appear frequently in specific documents. This latter frequency model was the choice for our analysis.

Despite "word" being part of the "bag-of-words" model's name, the features within a BOW do not necessarily need to be single words. One can opt for features either on word (*word n-grams*) or character (*character n-grams*) level. The *n* in n-gram refers to the length of the word or character sequence that represents a feature. For instance, the character bigrams at the beginning of this sentence are "Fo", "or", "r_", "_i", "in", etc. In our study, we chose to focus on such character bigrams, as they enable a nuanced observation of scribal variation and are known to be effective in handling data with inconsistencies (Eder 2013).[15] Additionally, we incorporated a specific criterion into our bigram selection: all of them need to include at least one brevigraph. Brevigraphs, glyphs used to denote abbreviations, are distinctive choices made by scribes. By concentrating on bigrams that contain a brevigraph, we can more directly investigate the scribe's individual style, while also reducing the reliance on the content of the texts themselves.

Constructing a BOW model based on elements like bigrams often results in a sparse frequency table (i.e., containing many 0-values). Therefore, it is common practice in stylometry to remove the least informative features from the model (Kestemont 2015). Typically, these are features with the lowest count or frequency. To manage this, we apply a frequency filter to the table, retaining only the 100 most frequent bigrams for our analysis. Keeping only the most frequent features offers an additional benefit: these features are likely present in all the texts, providing a stable basis for comparison across different manuscripts (Binongo 2003). Applying the frequency filter results in a compact computational representation of the text, whilst still being accurate. An important factor to still consider is the varying lengths of the texts in the corpus, which can significantly impact word frequencies (e.g., function words will have a higher (relative) frequency than content words in longer texts). To address this, we divided all texts into non-overlapping, consecutive segments of 5,000 characters each (see also Table 2). This segmentation ensures that longer and shorter texts can be compared reliably, without feature frequencies being influenced by text length (Karsdorp, Kestemont, and Allen 2021).

### Exploratory Scatterplots

Based on the BOW that represents the corpus, we created a scatterplot to visualise the stylistic distance between the production units of various scribes (Fig. 10). The scatterplot visualises each 5,000-characters segment as a dot within a two-dimensional space, determined by their stylistic similarities and dissimilarities. By representing our corpus in geometric terms, where each BOW column is a dimension or axis, we assign positions to texts based on their TF-IDF feature frequencies. Our BOW model initially features a configuration of 1,493 segments (rows) by 100 character-bigram features (columns). Using a technique called Principal Component Analysis (PCA), we compress these dimensions to 1,493x25, a transformation well-suited for textual data as evidenced by previous research (Kestemont 2012).[16] Subsequently, we employ UMap to further condense the model to 1,493x2 dimensions, thus enabling a two-dimensional scatterplot visualisation of our highly-dimensional textual data.[17]

---

[15] As mentioned earlier, our transcription process, while meticulously conducted, is not immune to errors. This is particularly relevant considering the use of HTR technology, which can introduce inaccuracies.

Observing the scatterplot reveals some compelling patterns. Most notably, the works of α (the Speculum Scribe) tend to cluster together towards the top of the plot, distinctly separate from other scribal oeuvres (Fig. 8). However, there are also some segments of α's work that appear more isolated, particularly those from the manuscript Vienna, ÖNB, SN 65, which is positioned farthest from α's main cluster. It is worth remembering that this manuscript also showed a notably low abbreviation density. This distinct separation within the scatterplot underlines the significant divergence in writing style found in Vienna, ÖNB, SN 65 compared to α's other works, raising the possibility that α may not have been the scribe responsible for these segments. The next section will further investigate the specific features that contribute to this difference in style and discuss what these findings imply.

Furthermore, we observe that Vienna, ÖNB, Cod. 12.857 (ca. 1375-1400) is positioned somewhat apart from the main cluster of α's segments (Fig. 10). Interestingly, it aligns with segments by scribe B from Brussels, KBR, 2979 (ca. 1350), and Saint Petersburg, BAN, O 256 (ca. 1325-1350). These manuscripts form a unique trio as they all contain Middle Dutch translations of the evangeliaria and were all located in Herne in the fourteenth century. While Brussels, KBR, 2979 was not transcribed by α, he significantly corrected the text in the manuscript. Notably, the text on folium 58r was erased, after which α penned an improved version (Kwakkel, 2002a, 224-225). Additionally, Kwakkel suggests that the Saint Petersburg manuscript could have served as the exemplar for the Viennese one (Kwakkel, 2002a, 47-52).[18] Considering our focus on character bigrams including brevigraphs, it is improbable that these manuscripts clustered due to similar content alone. Still, to verify whether Vienna, ÖNB, Cod. 12.857's divergent positioning is more style-driven than content-driven, we redid the analysis excluding the Saint Petersburg and Brussels manuscripts (and thus undoing any kind of influence these manuscripts have on the clustering). If Vienna, ÖNB, Cod. 12.857 then shifted closer to the main α-cluster, this would indicate that its initial proximity to the other manuscripts (in Fig. 10) was influenced by its content. However, Vienna, ÖNB, Cod. 12.857 remains isolated, suggesting a distinct writing style compared to α's other works. This isolation could mean that α closely replicated his exemplars in Vienna, ÖNB, Cod. 12.857, possibly indicating it as one of his earlier works where he adhered more strictly to his sources.

---

[16] https://scikit-learn.org/stable/modules/generated/sklearn.decomposition.PCA.html [accessed 22-12-2023]

[17] https://umap-learn.readthedocs.io/en/latest/ [accessed 22-12-2023]

[18] Kwakkel (2002a, p. 49) rules out Vienna, ÖNB, Cod. 12.857 as a copy of Brussels, KBR 2979, but our stylometric analysis reveals a strong similarity between both codices, specifically the segments by scribe B in the Brussels manuscript. Further analysis is warranted to explore whether α might have used both the Brussels and Saint Petersburg manuscripts as exemplars for his transcription of the Gospel of Matthew.

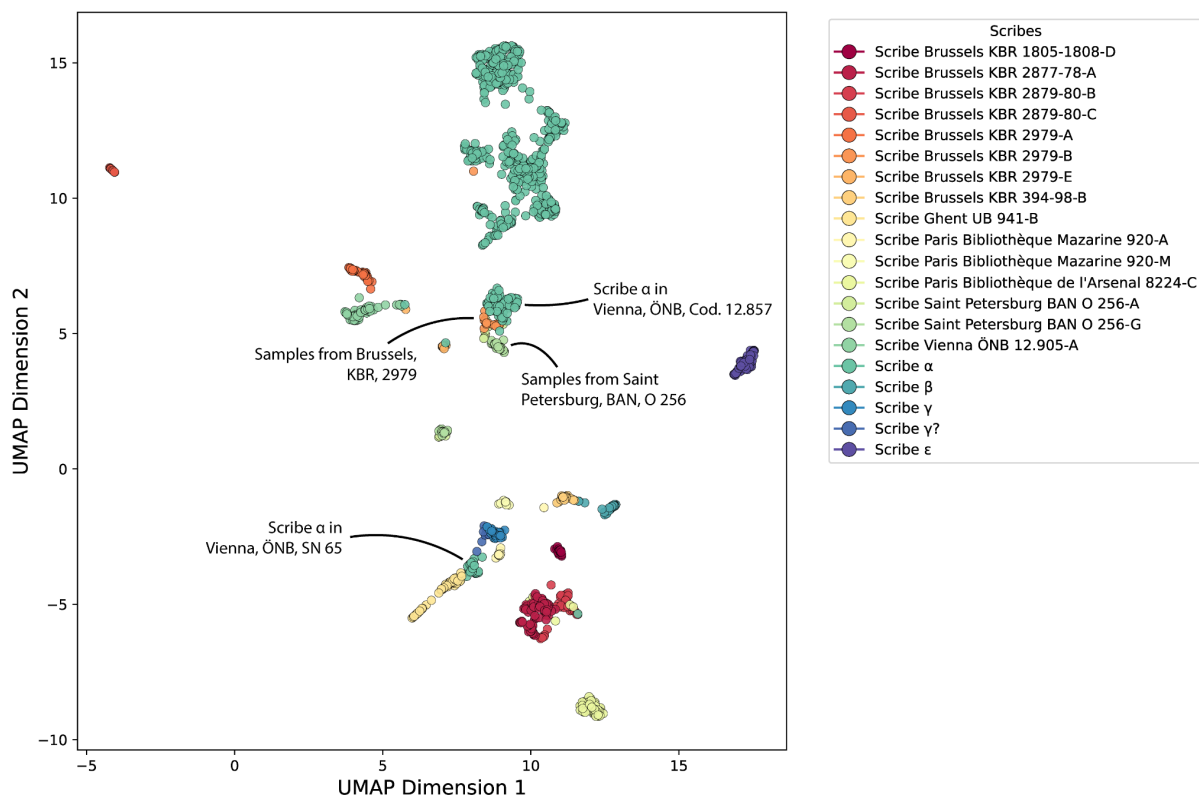

Fig. 10. Stylometric scatterplot of Middle Dutch manuscript scribes, using character-level bigrams with brevigraphs. This scatterplot, created using PCA and UMAP dimensionality reduction techniques, illustrates the stylistic variation among scribes in our corpus. Each dot represents a 5,000-character segment, coloured uniquely to correspond to a specific scribe (only scribes are retained that can contribute at least 10 segments). The spatial proximity of the dots indicates stylistic similarity.

We conducted our stylometric experiment once more, this time exclusively focusing on the works of α and β, who are recognized as the most prolific Herne scribes and known for their collaborative efforts (Fig. 11). Despite their intense collaborative scribal activities, the scatterplot distinctly separates their contributions. In the lower half, there is a small cluster that encompasses all of β's (the Necrology scribe's) Middle Dutch works. His works are isolated and clustered away from α's works. Additionally, the scatterplot shows Vienna, ÖNB, SN 65 in – again – an isolated position, reinforcing the notion that α may not have been its scribe. Similarly, the segments that make up Vienna, ÖNB, Cod. 12.857 are also situated distinctly away from the main cluster of α's works.

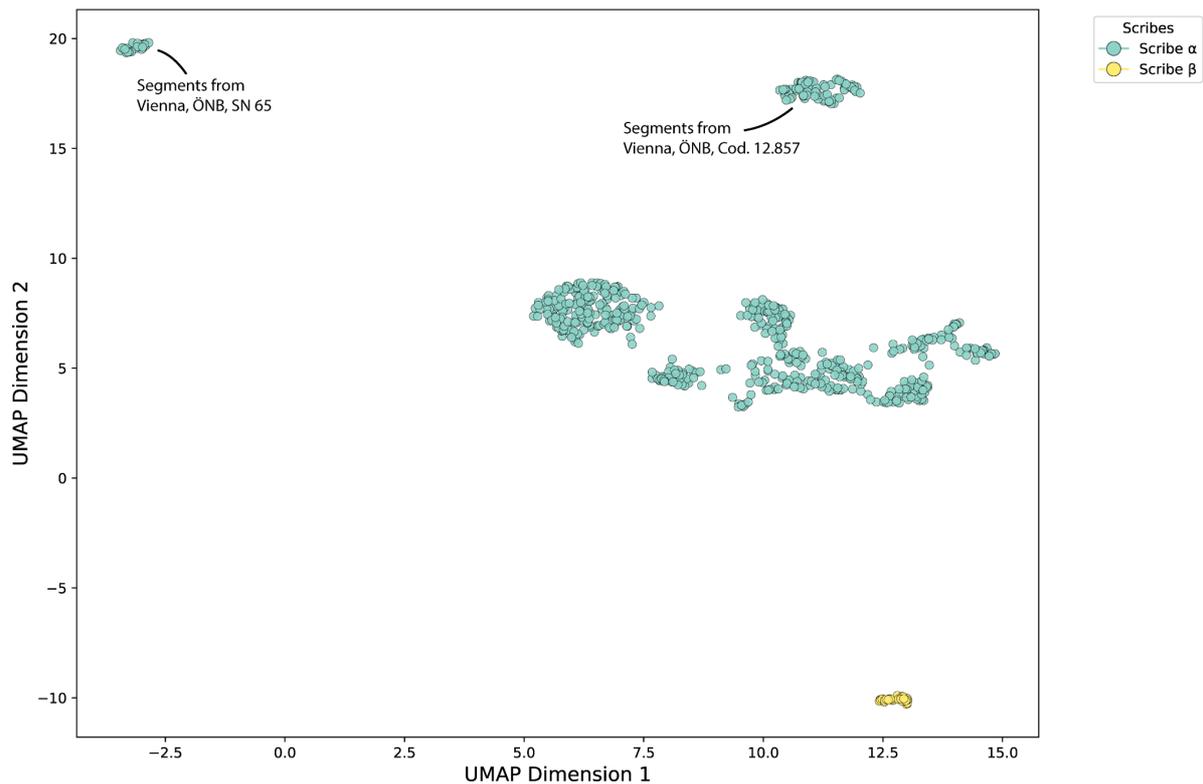

Fig. 11. Stylometric scatterplot of segments by scribes α and β. This graph displays each 5,000-character segment as a dot, with colour coding distinguishing segments attributed to either scribe α or β, as identified by Erik Kwakkel (2002a).

## Gamma (γ)

In our scatterplot analysis, we aim to explore Kwakkel's hypothesis that the scribe of Paris, BM, 920 production unit IV might be γ, who was also one of the scribes of Vienna, ÖNB, SN 65, contributing production units I, III and IV (Kwakkel 2002a, 257). Preliminary findings, evident in Figure 10, seem to corroborate this, showing a notable overlap in the works of γ and γ(?). However, a more cautious approach is essential. To mitigate the potential for bias, particularly from scribe α – who contributes a substantial number of segments to the analysis and hence could exert a kind of gravitational pull – we implement a more meticulous analysis strategy. This involves a downsampling technique, where we randomly select an equal number of 5,000-character segments (37, matching the number of segments available for scribe γ) from each Herne-associated scribe (i.e., α, β, ε, and γ). This approach effectively levels the playing field, allowing each scribe's style to be equally represented in the scatterplot and enabling a fair comparison. The final outcome of this analysis is revealing. In the resulting scatterplot (Fig. 12), we observe a close clustering of the segments belonging to γ and γ(?).

This clustering strongly suggests a significant stylistic overlap between the two, reinforcing Kwakkel's hypothesis that γ and γ(?) are one and the same scribe.[19]

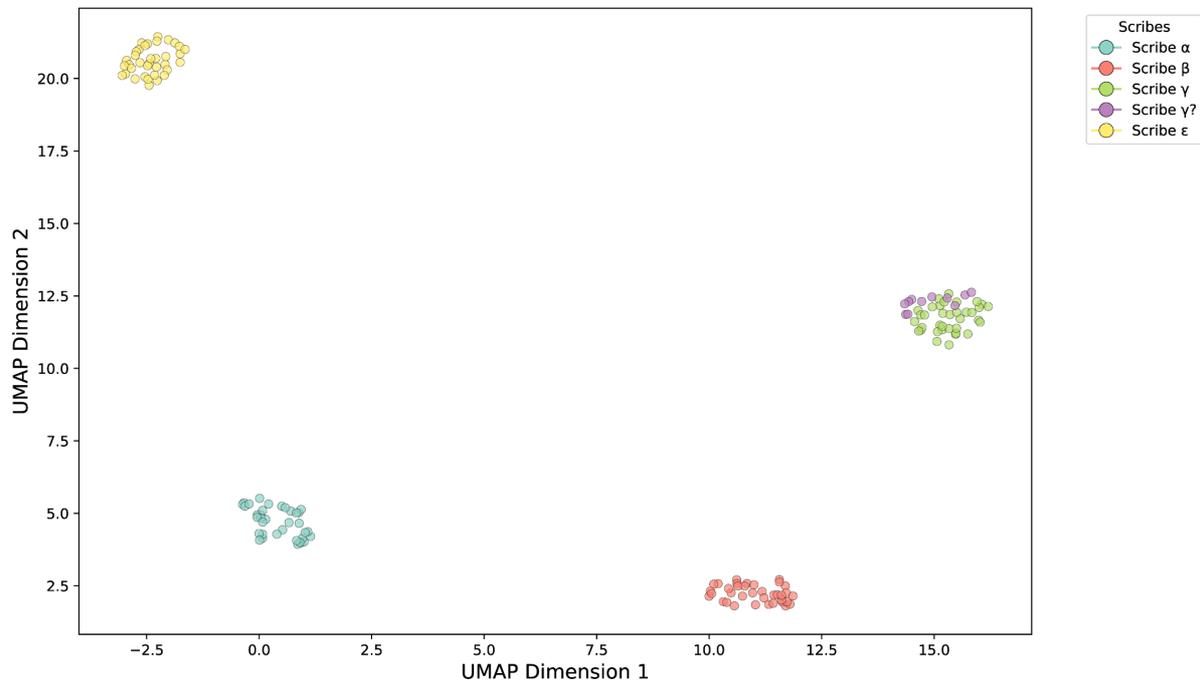

Fig. 12. Stylometric analysis of scribe γ and γ(?). This scatterplot illustrates the clustering of segments attributed to scribes γ and γ(?) after downsampling.

## Outlier Detection

In previous analyses, segments from Vienna, ÖNB, SN 65 (production unit II), attributed to scribe α, have often exhibited atypical behaviour. To further examine this anomaly, we conduct what is commonly called an 'outlier detection' analysis. This approach employs machine learning techniques to pinpoint key features that distinguish between various texts. To this purpose, we initially gather all 5,000-character segments accredited to scribe α. We then create a BOW model for these segments (a total of 706), focusing on the 100 most frequent character bigrams that include at least one brevigraph. Next, we apply a one-class support vector machine (OneClassSVM) model to our dataset.[20] An SVM is a type of machine learning

---

[19] This finding is further corroborated by a nearest neighbour classifier, as detailed in the online code (https://github.com/Caroline-Vandyck/making-characters-count). In a classification setup, γ(?) is definitively attributed to γ. Additionally, the abbreviation densities for γ and γ(?) are notably similar, at 0.28 and 0.26 respectively.

[20] In our set-up, the model is trained in a leave-one-out fashion. A OneClassSVM is trained on the BOW representation of all segments from α's production units, excluding the segments from the current unit under analysis. Once trained, the model is then applied to the excluded production unit. This tests if the segments in this unit conform to the learned style (inliers) or deviate from it (outliers). For each unit, the proportion of

algorithm that is particularly good at identifying outliers in a dataset. In our context, it is trained on the bigram features in our BOW model, allowing it to learn the 'normal' pattern of α's writing style. Once trained, the SVM model is used to predict whether a segment is an 'inlier' (conforming to the typical style of α) or an 'outlier' (deviating from α's usual style). The results of this model show that all segments from Vienna, ÖNB, SN 65 are labelled as outliers (Fig. 13), further substantiating its atypical nature in the context of α's work.

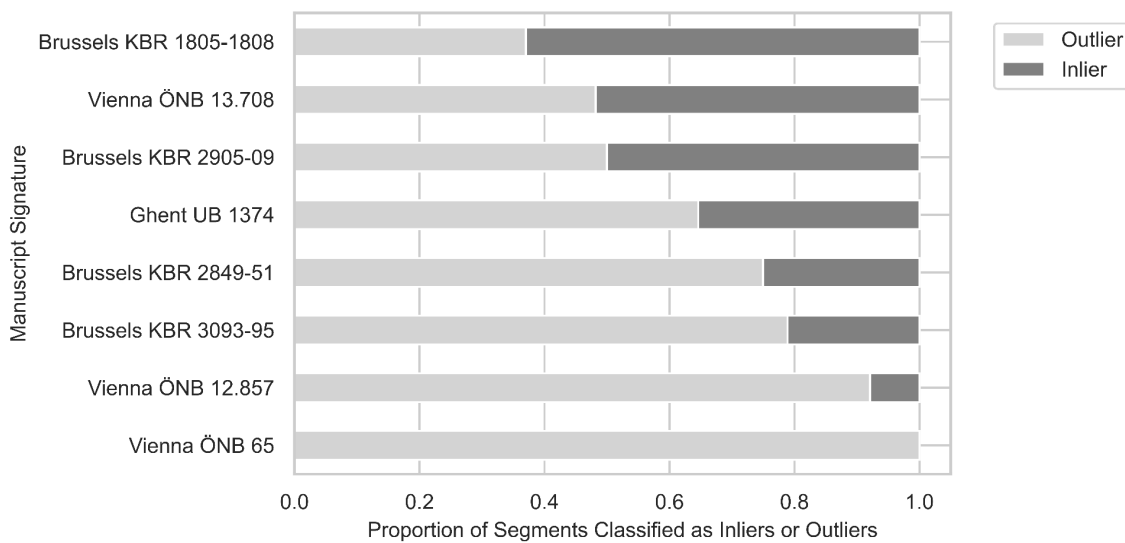

Fig. 13. Stacked bar chart illustrating the proportion of inliers and outliers in each manuscript signature attributed to scribe α. The analysis employs a OneClassSVM model trained in a leave-one-out fashion on a BOW representation of character bigrams including brevigraphs. Each bar represents a different manuscript signature, with the percentage of segments classified as outliers and inliers.

With the identification of α's segments from Vienna, ÖNB, SN 65 as outliers, a critical question emerges: what specific features cause a distinction between this manuscript and α's other works? To answer this, we employ a Random Forest Classifier to analyse feature importances.[21] This model, again, trained on BOW representation of the segments, detects patterns unique to the Viennese manuscript. After training, the model assesses each feature's 'power' in differentiating between two classes: α's segments from Vienna, ÖNB, SN 65, and those from other manuscripts.

---

segments classified as inliers (1) and outliers (-1) is calculated. This approach ensures that the model learns the 'normal' writing style of α, excluding potential stylistic variations present in the unit being tested. Finally, the data is aggregated at the codex level to analyse the overall consistency of α's writing across different codices.

[21] The Random Forest Classifier is adept at handling large datasets and providing insights into the significance of various features in prediction. It builds decision trees by splitting nodes based on feature values, and Mean Decrease in Impurity (MDI) is calculated for each feature. MDI quantifies a feature's contribution to the homogenization of nodes within trees. Features with higher MDI values are more crucial in dividing the data into respective classes.

Our analysis reveals significant differences in abbreviation usage in Vienna, ÖNB, SN 65 (production unit II) compared to α's other works (Fig. 14). Notably, the glyphs 'ē', 'ī' and 'ĵ', as well as the abbreviation 'nᵗ', characteristic of scribe α's style, are almost entirely absent in the Viennese codex. Moreover, a paleographic inspection of the script in Vienna, ÖNB, SN 65, coupled with a comparison to α's typical writing style in other manuscripts, reinforces our impression that we are dealing with different scribes.[22]

In summary, it has become evident that scribe α was not the author of production unit II in the Vienna, ÖNB, SN 65 manuscript. This discovery raises the possibility that the manuscript may not have been produced in Herne at all, since it also lacks the correctional delta mark, typical of the Herne Carhusians' scribal practice. This finding also has significant implications for scribe γ, who is another contributor to this manuscript. The hypothesis of γ being a Carthusian monk from Herne now becomes less certain since it was primarily based on his supposed collaboration with α in the creation of Vienna, ÖNB, SN 65. With the new understanding that α did not contribute to this manuscript, γ's residency in Herne becomes questionable, especially considering his lack of collaboration with other scribes. Even though γ was responsible for production unit IV of Paris, BM, 920, this does not determine γ's residency. In this manuscript, α's involvement is limited to corrections and text additions, which does not necessarily imply a Herne origin for the manuscript. Consequently, both the origin of the Vienna, ÖNB, SN 65 manuscript and the residency of scribe γ remain unresolved and open to further investigation.

---

[22] In a recent personal communication [24/10/2023], Kwakkel acknowledged that, upon re-examination, the script of Vienna, ÖNB, SN 65 (production unit II) indeed shows discernible differences from the handwriting in other works ascribed to scribe α. Although there are similarities, the overall appearance of the two hands is distinctly different.

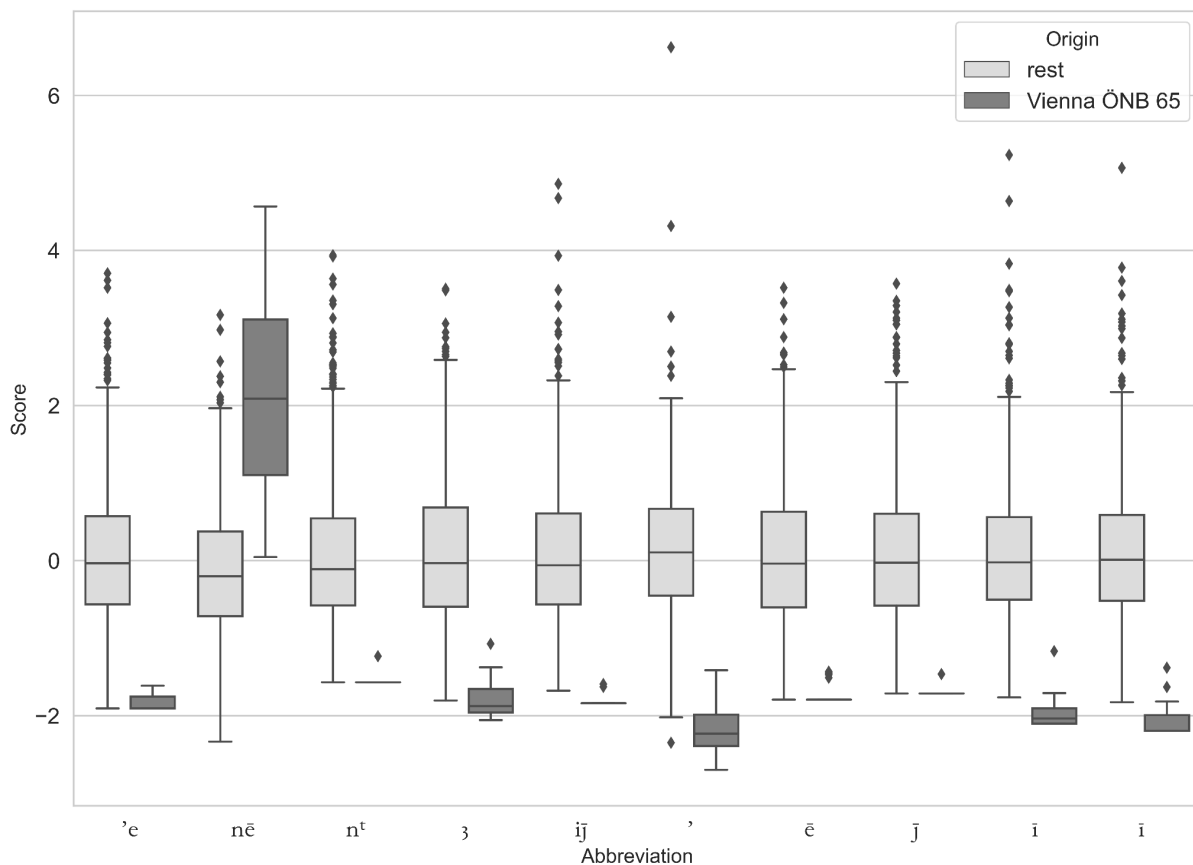

Fig. 14. Boxplots with the most important features that help to distinguish between different text segments (Vienna, ÖNB, SN 65 vs. the rest).

## Conclusion

Abbreviations, while often overlooked as minor elements of medieval scribes' writing styles, have shown immense power in the case of the Herne scribes. The extensive use of glyphs for abbreviating letter sequences by these 14th-century scribes offers a unique stylistic window into their scribal activities. By calculating abbreviation densities, we demonstrated that the abbreviation usage of various Herne scribes ranks among the most dense in medieval vernacular languages. Simultaneously, examining abbreviation densities at the manuscript level has lead to intriguing findings, where outliers became evident. The real value of our stylometric study in this work is twofold: not only does it involve a thorough examination of abbreviation densities, but it also innovatively employs a Bag-of-Words model populated with character bigrams that include at least one brevigraph as features. This carefully considered method allowed us to either support or challenge hypotheses from Kwakkel's seminal thesis (2002a). While Kwakkel's approach is largely rooted in paleography and codicology, our study at the linguistic level offered a valuable complementary perspective. For instance, we affirmed Kwakkel's paleographic observation that production unit IV of Paris, BM, 920 was indeed written by scribe γ, who also contributed to production units I, III, and IV in Vienna ÖNB, SN

65. However, we disproved his assumption that scribe α contributed to production unit II of Vienna ÖNB, SN 65. This manuscript diverges stylistically from α's other writings in several respects. This discovery triggered a domino effect, leading to the conclusion that not only the manuscript's provenance is disconnected from Herne, but also the residency of scribe γ is now uncertain. All these revelations were concealed within some of the smallest elements of written language. When handled with care (and the grapheme library), small elements like brevigraphs prove to be a great source of information.

## References

Biemans, J.A.A.M. 1997. *Onsen Speghele Ystoriale in Vlaemsche. Codicologisch Onderzoek Naar de Overle-vering van de «Spiegel Historiael» van Jacob van Maerlant. Philip Utenbroeke En Lodewijk van Velthem, Niet Een Beschrijving van de Handschriften En Fragmenten.* 2 vols. Leuven: Peeters Publishers.

Binongo, J. 2003. "Who Wrote the 15th Book of Oz? An Application of Multivariate Analysis to Authorship Attribution." *Chance* 16 (2): 9–17.

Dalen-Oskam, Karina van. 2007. "Kwantificeren van Stijl." *Tijdschrift Voor Nederlandse Taal-En Letterkunde* 123: 37–54.

Eder, M. 2013. "Mind Your Corpus: Systematic Errors in Authorship Attribution." *Literary and Linguistic Computing* 28 (4): 604–14.
Gabriël, R., & Kestemont, M. (2023). T*he Heber-Serrure Codex (Ghent, University Library, Ms. 1374)* (0.0-rc) [Data set]. Zenodo. https://doi.org/10.5281/zenodo.8397029

Gaens, Tom, and Jan de Grauwe. 2006. *De Kracht van de Stilte: Geest & Geschiedenis van de Kartuizerorde.* Leuven: Peeters.

Guigues. 2001. *Coutumes de Chartreuse.* Edited by Maurice Laporte. Réimpr. de la 1re éd. rev. et corr. Sources Chrétiennes 313. Paris: Les Éd. du Cerf.

Haverals, Wouter, and Mike Kestemont. 2023a. "From Exemplar to Copy: The Scribal Appropriation of a Hadewijch Manuscript Computationally Explored." *Journal of Data Mining and Digital Humanities*, 1–21. https://doi.org/10.46298/JDMDH.10206.

———. 2023b. "The Middle Dutch Manuscripts Surviving from the Carthusian Monastery of Herne (14th Century): Constructing an Open Dataset of Digital Transcriptions." In . Paris.

Honkapohja, Alpo. 2021. "Digital Approaches to Manuscript Abbreviations: Where Are We at the Beginning of the 2020s?" *Digital Medievalist* 14 (1). https://doi.org/10.16995/dm.88.

Karsdorp, Folgert, Mike Kestemont, and Riddell Allen. 2021. *Humanities Data Analysis: Case Studies with Python.* Princeton University Press.

Kestemont, Mike. 2012. "Stylometry for Medieval Authorship Studies: An Application to Rhyme Words." *Digital Philology. A Journal of Medieval Cultures* 1: 42–72.

———. 2015. "A Computational Analysis of the Scribal Profiles on Two of the Oldest Manuscripts of Hadewijch's 'Letters.'" *Scriptorium* 69 (2): 159–77.

———. 2018. "Stylometric Authorship Attribution for the Middle Dutch Mystical Tradition from Groenendaal." *Journal of the Low Countries Studies* 42 (3): 203–37. https://doi.org/10.1080/03096564.2016.1252077.

Kestemont, Mike, Walter Daelemans, and Guy De Pauw. 2010. "Weigh Your Words— Memory-Based Lemmatization for Middle Dutch." *Literary and Linguistic Computing* 25 (3): 287–301. https://doi.org/10.1093/llc/fqq011.

Kestemont, Mike, and Karina van Dalen-Oskam. 2009. "Predicting the Past: Memory Based Copyist and Author Discrimination in Medieval Epics." In *Proceedings of the Twenty-First Benelux Conference on Artificial Intelligence*, 121–28. Eindhoven.

Kwakkel, Erik. 2002a. *Die Dietsche Boeke Die Ons Toebehoeren de Kartuizers van Herne En de Productie van Middelnederlandse Handschriften in de Regio Brussel (1350-1400)*. Miscellanea Neerlandica, XXVII. Leuven: Peeters.

———. 2002b. "Towards a Terminology for the Analysis of Composite Manuscripts." *Gazette Du Livre Médiéval* 41: 12–19.

———. 2003. "A Meadow without Flowers. What Happened to the Middle Dutch Manuscripts from the Charterhouse Herne?" *Quærendo* 33: 191–211.

———. 2018. *Books before Print*. Medieval Media Cultures. Leeds: ARC Humanities Press.

Lievens, Robert. 1960. "In margine. De dichter Hein van Aken." *Spiegel der Letteren* 4 (1): 57–74.

McIntosh, Angus. 1975. "Scribal Profiles from Middle English Texts." *Neuphilologische Mitteilungen* 76 (2): 218–35.

Mierlo, Jozef van. 1929. "Arnout en Willem." *Verslagen en Mededelingen van de Koninklijke Vlaamse Academie Voor Taal- En Letterkunde*, 757–88.

Robinson, P., & Solopova, E. (1993). "Guidelines for Transcription of the Manuscripts of the Wife of Bath's Prologue." *The Canterbury Tales Project Occasional Papers*, 1, 19-52.

Sebastiani, F. (2001). "Machine Learning in Automated Text Categorization." ArXiv, cs.IR/0110053. [Online]. Available: https://api.semanticscholar.org/CorpusID:3091

Stutzmann, Dominique. 2011. "Paléographie Statistique Pour Décrire, Identifier, Dater... Normaliser Pour Coopérer et Aller plus Loin?" In *Kodikologie Und Paläographie Im Digitalen Zeitalter 2 - Codicology and Palaeography in the Digital Age 2*, by Franz Fischer, Christiane Fritze, and Georg Vogeler, 247–77. Books on Demand.

Thaisen, Jacob. 2010. "Probabilistic Modeling of Middle English Scribal Orthographies." In *Digital Humanities 2010. Conference Abstracts*, 37–39. London.